\documentclass[twocolumn]{aa}
\usepackage{epsfig}
\def\simlt{\ \raise -2.truept\hbox{\rlap{\hbox{$\sim$}}\raise5.truept   %
\hbox{$<$}\ }}
\def\simgt{\ \raise -2.truept\hbox{\rlap{\hbox{$\sim$}}\raise5.truept   %
\hbox{$>$}\ }}                                                          %

\def\be{\begin{equation}}
\def\ee{\end{equation}}
\def\newline{\hfil\break}

\def\la{\mathrel{\hbox{\rlap{\hbox{\lower4pt\hbox{$\sim$}}}\hbox{$<$}}}}
\def\ga{\mathrel{\hbox{\rlap{\hbox{\lower4pt\hbox{$\sim$}}}\hbox{$>$}}}}

\begin{document}
\title{On the ICS interpretation of the Hard X-Ray Excesses in Galaxy Clusters:
the case of Ophiuchus}
   \author{S. Colafrancesco\inst{1,2} and P. Marchegiani\inst{2,3}}
   \offprints{S. Colafrancesco}
\institute{   ASI-ASDC
              c/o ESRIN, Via G. Galilei snc, I-00040 Frascati, Italy
              Email: sergio.colafrancesco@asi.it
 \and
              INAF - Osservatorio Astronomico di Roma
              via Frascati 33, I-00040 Monteporzio, Italy.
              Email: cola@mporzio.astro.it
 \and
              Dipartimento di Fisica, Universit\`a di Roma La Sapienza, P.le A. Moro 2, Roma, Italy
              Email: marchegiani@mporzio.astro.it
             }
\date{Received 13 February 2009 / Accepted 14 April 2009 }
\authorrunning {S. Colafrancesco and P. Marchegiani}
\titlerunning {Consequences of HXR excesses in Clusters}
\abstract
   {High energy electrons can produce Hard X-Ray (HXR) emission
   in galaxy clusters via Inverse Compton Scattering (ICS) of CMB photons.}
   {
   We discuss here the consequences of the presence of such high energy
   particles for the multi-frequency emissivity of the same clusters and for the
   structure of their atmospheres. }
  {
  We derive predictions for the ICS HXR emission in the specific case of the Ophiuchus
  cluster under three main scenarios for producing high-E electrons:
  primary cosmic ray model, secondary cosmic rays model and neutralino DM annihilation
  scenario. We further discuss the predictions of the Warming Ray model
  for the cluster atmosphere.
  Under the assumption to fit the HXR emission observed in Ophiuchus, we explore the
  consequences that these electron populations induce on the cluster atmosphere.}
   {
  We find that:
  i) primary electrons can be marginally consistent with the available data
  provided that the electron spectrum is cutoff at $E \simlt 30$ ($90$) MeV
  for electron spectral index values of 3.5 (4.4);
  ii) secondary electron models from pp collisions are inconsistent with the
  viable gamma-ray limits, cosmic ray protons produce too much heating of
  the intra cluster (IC) gas and their pressure at the cluster center largely
  exceeds the thermal one;
  iii) secondary electron models from DM annihilation are inconsistent with the
  gamma-ray and radio limits, and produce too much heating of the IC gas at the
  cluster center, unless the neutralino annihilation cross section is much lower
  than the proposed value.
  In that case, however, such models no longer reproduce the HXR
  excess in Ophiuchus.}
  {
  We conclude that ICS by secondary electrons from both neutralino DM annihilation and
  pp collisions cannot be the mechanism responsible for the HXR excess emission;
  primary electrons are still a marginally viable solution provided that their spectrum has a
  low-energy cutoff at $E \simlt 30-90$ MeV.
  We also find that diffuse radio emission localized at the cluster center is expected
  in all these models and requires quite low values of the average magnetic field
  ($B\sim0.1-0.2$ $\mu$G in primary and secondary-pp models; $B\sim0.055-0.39$ $\mu$G in secondary-DM models)
  to agree with the observations.
  Finally, the WR model (with $B\sim0.4-2.0$ $\mu$G) offers, so far, the best description of the cluster in
  terms of the temperature distribution, heating and pressure and multi-frequency spectral
  energy distribution.
  Fermi observations of Ophiuchus will set further constraints to this model.}

 \keywords{Cosmology; Galaxies: clusters: theory; Dark Matter, Cosmic Rays}
 \maketitle

\section{Introduction}
 \label{sec.intro}

Hard X-Ray (HXR) excess emission in galaxy clusters has been
observed in the direction of several nearby systems (see
\cite{Nevalainen2004}) but its origin is still disputed.
It has been proposed that such HXR emission is due to inverse
Compton scattering (ICS) of relativistic electrons with the cosmic
microwave background (CMB) (Blasi \& Colafrancesco 1999, Atoyan \&
Volk 2000, Ensslin \& Biermann 1998, Sarazin 1999, Brunetti 2004,
Profumo 2008, see Petrosian et al. 2008 for a recent review), to
bremsstrahlung emission from a supra-thermal electron population
(Dogiel et al. 2007, see Petrosian et al. 2008 for a recent
review) or to a population of PeV electrons that would radiate in
hard X-rays through synchrotron emission (Timokhin, Aharonian \&
Neronov 2004; Inoue, Aharonian \& Sugiyama 2005). None of these
models has been definitely proven or rejected, so far, due to the
lack of instrumental sensitivity (spatial and spectral) of the
available experiments operating in the HXR band.

In such a context, the Ophiuchus cluster (z=0.028, Johnston et al.
1981) has been recently at the center of an interesting dispute
concerning the combination of new observational evidence for the
HXR emission and various theoretical considerations on its
origin.\\
The Ophiuchus cluster seems to have a high plasma temperature $kT
\sim 10$ keV (Johnston et al. 1981).
Measurements of the IC gas temperature vary from $8.5 \pm 0.5$ keV
(INTEGRAL; Eckert et al. 2008) up to $9.5^{+1.4} _{-1.1}$ keV
(Swift/BAT; Ajello et al. 2009). Watanabe et al. (2001) also found
a large ($20' \times 30'$), hot ($kT > 13$ keV) region, 20 arcmin
west of the cluster center, from which they concluded that the
cluster is not dynamically relaxed, and suggested that it
experienced a major merging event in the recent past ($t \simlt 1$
Gyr).\\
Eckert et al. (2008) have recently reported a tentatively resolved
($\sim 5'$) X-ray source at the cluster center, and indicated the
presence of a non-thermal emission tail with a flux $(10.1 \pm
2.5) \cdot 10^{-12}$ erg cm$^{-2}$ s$^{-1 }$ in the 20--60 keV
energy band. These authors interpreted the non-thermal hard X-ray
emission as due to ICS emission from relativistic electrons
scattered off the CMB in the intra-cluster (IC) medium.
Suzaku observations of the Ophiuchus cluster by Fujita et al.
(2008) have, however, failed to detect the non-thermal component
detected by Eckert et al. (2008), although their quoted upper
limit of $2.8 \cdot 10^{-11}$ erg cm$^{-2}$ s$^{-1 }$ in the
20--60 keV energy band is still compatible with the INTEGRAL
detection. Ajello et al. (2009) have found, using Swift/BAT
spectra (with a IC gas temperature of $kT = 9.5$ keV), an upper
limit on the Ophiuchus non-thermal X-ray emission in the 20--60
keV band, of $7.2 \cdot 10^{-12}$ erg cm$^{-2}$ s$^{-1}$ ($90 \%$
c.l.).
We notice that the INTEGRAL detection and the Swift-BAT upper
limit are consistent, at the same $90 \%$ confidence level, in the
flux range $(6.1 - 7.2)\cdot 10^{-12}$ erg cm$^{-2}$ s$^{-1}$.
This is the flux range in which the HXR excess detected from
Ophiuchus is consistent with both Swift-BAT and INTEGRAL
observations. In our study of the origin of such HXR excess, we
refer to the value $F_{20-60 {\rm keV}} =7.2 \cdot 10^{-12}$ erg
cm$^{-2}$ s$^{-1}$ as the maximum value of its flux and we discuss
how our results change by considering also the minimum flux of the
HXR, $F_{20-60 {\rm keV}} =6.1 \cdot 10^{-12}$ erg cm$^{-2}$
s$^{-1}$, that is $\sim 15 \%$ less than the previous maximum
value.

The possible presence of an ICS tail of HXR emission in Ophiuchus
was previously related to the identification of the steep-spectrum
radio source MSH 17-203 (also dubbed Cul 1709-231) as a radio mini
halo (Johnston et al. 1981), that would thus imply the presence of
relativistic electrons, and hence the relative ICS emission
emerging from the thermal bremsstrahlung emission in the X-ray
band at $E \simgt 20$ keV.\\
However, a recent high resolution study (Perez-Torres et al. 2008)
made with 240 and 607 MHz GMRT radio observations of the Ophiuchus
cluster of galaxies, along with archival 74 and 1400 MHz VLA data,
indicates that there is no significant diffuse radio emission in
the core of the Ophiuchus, and that the previous measurements of
radio flux from the MSH 17-203 source (Slee \& Higgins 1975, Slee
1977) do not refer to the radio halo of the cluster: these authors
present new upper limits to the integrated, diffuse non-thermal
radio emission from the cluster core.
More recently, Govoni et al. (2009) pointed out that there is
indeed a diffuse radio mini-halo located at the center of the
Ophiuchus cluster, with angular size of $\sim 9 \times 12$
arcmin$^2$; the flux of the mini halo is $\sim 8$ times lower than
the old measurement of Johnston et al. (1981), which had a
resolution of 80 arcmin (see details in Perez-Torres et al. 2008).

There is no other information on non-thermal emission from
Ophiuchus: gamma-ray emission from this cluster has been not
detected and therefore the only information we have directly on
the high-E particle population of Ophiuchus is an upper limit
obtained by EGRET $F(>100 \mbox{ MeV}) = 5 \cdot 10^{-8}$
cm$^{-2}$ s$^{-1}$ (Reimer et al. 2003).

In such an obervational scenario (that is similar to other
clusters where an HXR emission detection has been claimed)  the
HXR emission excess from Ophiuchus has been recently interpreted
as ICS emission from either a population of primary cosmic ray
electrons (Eckert et al. 2008) or secondary electrons produced in
neutralino DM annihilation (Profumo 2008). In particular, Profumo
(2008) proposed that a combination of three different neutralino
DM models [$M_{\chi}= 81 (W^+W^-), 40(b {\bar b}$) and $10 (\tau^+
\tau^-)$ GeV] is consistent with all non-thermal emission data for
Ophiuchus, from radio to HXR and gamma-rays.
The available data on diffuse radio emission in the core of
Ophiuchus and the overall analysis of its multi-frequency SED
further led Perez-Torres et al. (2008) to conclude that i) a
synchrotron+ICS model from primary cosmic ray electrons is in
marginal agreement with the the available data, with a range of
magnetic field values $B\sim 0.02-0.3$ $\mu$G; ii) that a pure
neutralino annihilation scenario cannot reproduce both radio and
HXR emission, unless extremely low magnetic field values
($10^{-2}$ to $10^{-3}$ $\mu$G) are assumed; iii) a scenario in
which synchrotron and ICS arise from PeV electron-positron pairs
(via interactions with the CMB), can also be ruled out, as it
predicts a non-thermal soft X-ray emission that largely exceeds
the thermal bremsstrahlung emission measured by INTEGRAL.

In this paper we take a more radical approach to the problem of
the HXR emission of Ophiuchus and we consider not only the SED
properties of synchrotron plus ICS scenarios (from both primary
and secondary electrons) but also the physical consequences of the
ICS origin of the HXR emission in all models so far viable:
primary electron model (Sect 2.1), secondary electron models from
pp collisions (Sect.2.2) and from DM annihilation (Sect.2.3) and
finally a Warming Ray model (Sect.3). We will discuss our
conclusions in Sect.4.

Throughout the paper, we use a flat, vacuum--dominated cosmological model with
$\Omega_m = 0.3$, $\Omega_{\Lambda} = 0.7$ and $h = 0.7$.

\section{Modeling the electron populations in Ophiuchus}
 \label{sec.model}

The spatial distribution of the intra cluster (IC) thermal plasma
in the Ophiuchus cluster can be represented by:
\begin{equation}
n_{th}(r)=n_{th,0} \left[ 1+\left( \frac{r}{r_c} \right)^2 \right]^{-q_{th}}
 \label{eq.nth}
 \end{equation}
with $r_c= 108 h_{70}^{-1}$ kpc and $q_{th}=0.96$ (Watanabe et al.
2001), and $n_{th,0}=1.77 \times 10^{-2} h_{70}^{1/2}$ cm$^{-3}$
(Johnston et al. 1981). The available data for this cluster
indicate, out to a reasonable extent, an isothermal temperature
profile with a temperature $kT \sim 9.9$ keV (Watanabe et al.
2001, Ajello et al. 2009); the virial radius of the cluster is
$R\sim 1.7 h_{70}^{-1}$ Mpc (Mohr et al. 1999).

Non-thermal electrons that can be able to produce the cluster HXR emission by
ICS on CMB photons must have energies $E_e \approx 0.35 \mbox{ GeV}
(E/\mbox{keV})^{1/2}$ in the range $\approx 1.6-2.7$ GeV, in the case of
the HXR emission observed in the 20--60 keV range (see, e.g., Colafrancesco et
al. 2005).

In this section we discuss the predictions of various models for the origin of
the high-E electrons: i) primary electron model (PEM) (see, e.g.,
Colafrancesco, Marchegiani \& Perola 2005 and references therein); ii)
secondary electron model produced by proton-proton (pp) collisions in the
cluster atmosphere (SEM-pp) (see, e.g., Colafrancesco \& Blasi 1998, Blasi \&
Colafrancesco 1999, Marchegiani et al. 2007); iii) secondary electron models
produced by neutralino Dark Matter annihilation (SEM-DM) (see e.g.
Colafrancesco et al. 2006, Profumo 2008).\\
The electron spectra expected in the previous models are
normalized by assuming that the produced ICS HXR emission equals
the Swift-BAT/INTEGRAL data.

In addition to the previous models, we also consider in the next
Sect.3 a self-consistent warming-ray (WR) electron model (see e.g
Colafrancesco \& Marchegiani 2008; Colafrancesco et al. 2004) in
which the cluster atmosphere is heated - in a quasi stationary
equilibrium condition between heating and cooling - by the
interactions of non-thermal cosmic-ray protons with the IC gas.
Note that this WR model reproduces the X-ray properties of the
thermal IC gas (namely its temperature and density profiles) and,
therefore, we use this constraint to predict the cluster ICS HXR
emission.

\subsection{Primary Electron Model (PEM)}

\begin{figure}[ht]
\begin{center}
\vbox{
 \epsfig{file=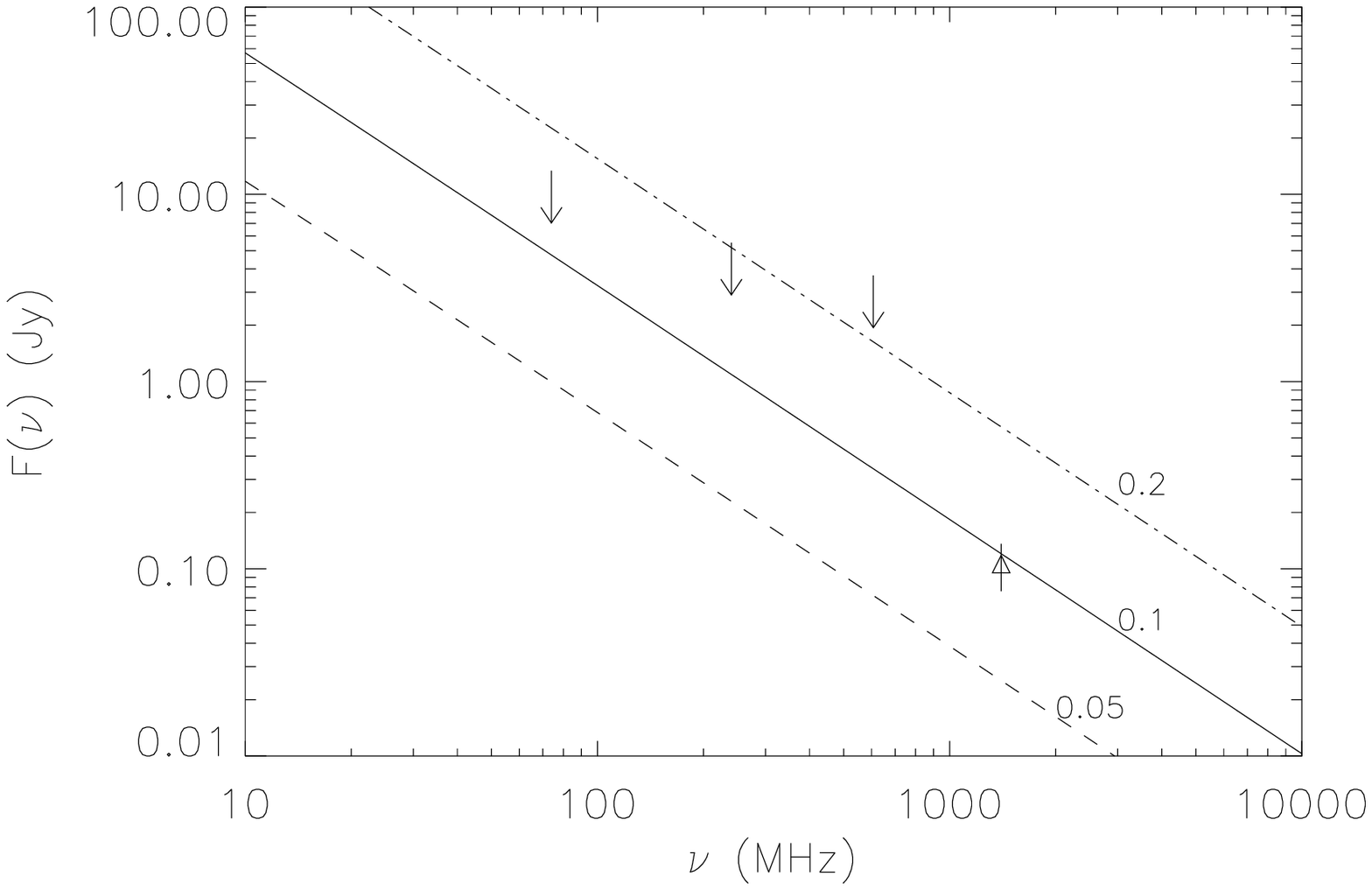,height=8.cm,width=8.cm,angle=0.0}
 \epsfig{file=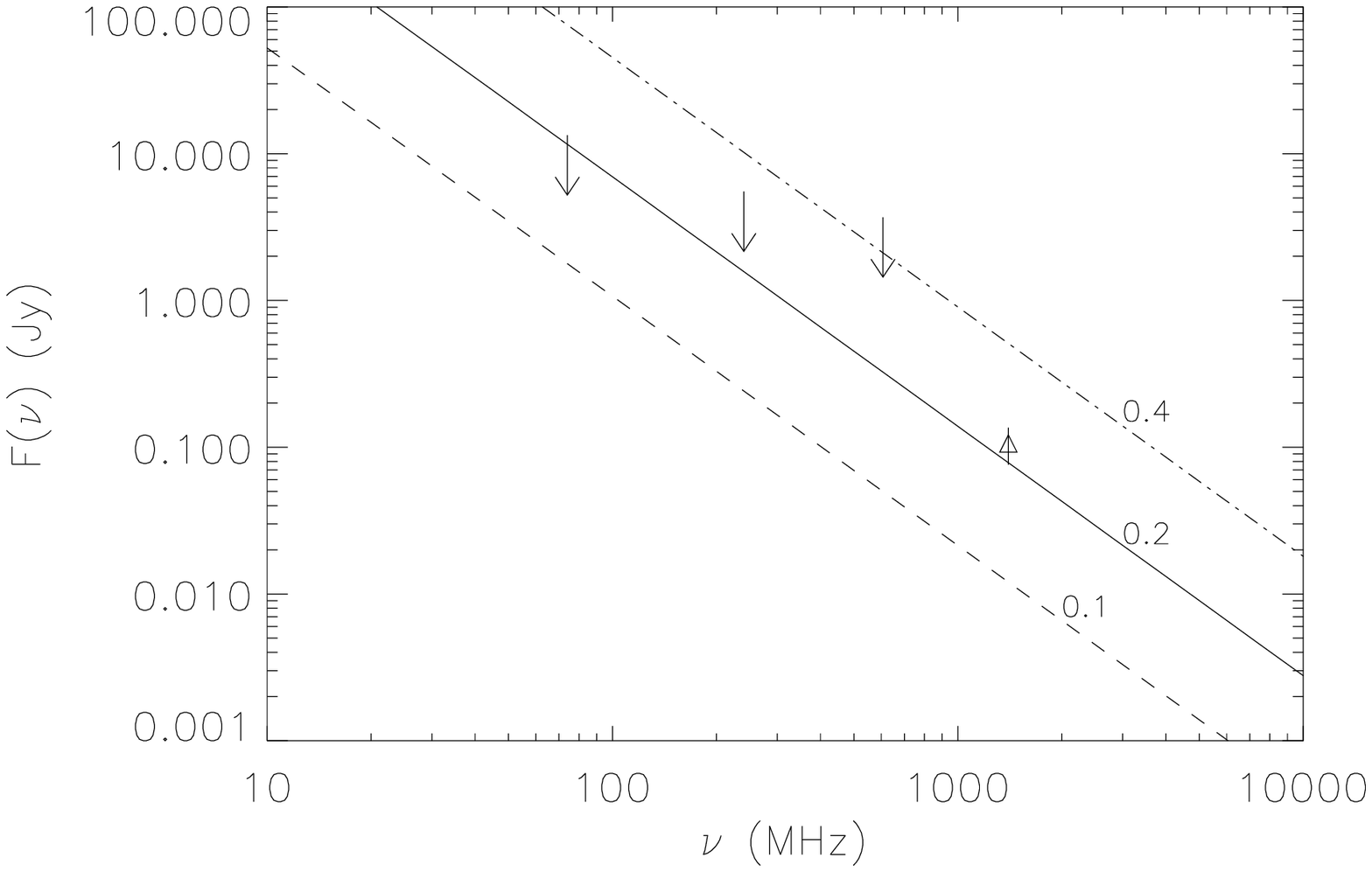,height=8.cm,width=8.cm,angle=0.0}
}
\end{center}
 \caption{\footnotesize{The diffuse radio emission spectrum as produced
 from primary electrons with $p=3.5$ (upper panel) and $p=4.4$ (lower panel)
 is shown for different values of the uniform magnetic
 field (in units of $\mu$G), as labelled.
Data are from Perez-Torres et al. (2008) (upper limits) and Govoni
et al. (2009) (point at 1.4 GHz).
 }}
 \label{fig.radioprim}
\end{figure}

The high-E electron spectrum is usually best constrained by using
the radio halo synchrotron spectrum that provides direct
information on the electron spectral shape (see e.g. Colafrancesco
et al. 2005 for a discussion).
However, the upper limits on the Ophiuchus radio halo (see
Perez-Torres et al. 2008) and the measurement at 1.4 GHz made by
Govoni et al. (2009) are not sufficient to determine precisely the
electron spectrum.  Therefore, we choose to adopt here a simple
power-law model
\begin{equation}
N_e(E,r)=N_{e,0} (E/\mbox{GeV})^{-p} \cdot g(r) \;
 \label{spettroele}
\end{equation}
with $p=3.5$, which corresponds to a radio spectral index of
$\alpha_R=1.25$, typical of radio halos in galaxy clusters and
consistent with the available radio data on Ophiuchus; we also
consider the case with $p=4.4$, which provides $\alpha_R=1.7$,
which is, at 1$\sigma$ level, the maximum value of the spectral
index allowed by radio data (see, e.g., Fig.\ref{fig.radioprim}).
The radial function $g(r)$ should be constrained, in principle, by
measuring the radial shape of the ICS emission of the relativistic
electrons (the radio measures are not sufficient to constrain this
shape; see discussion in Colafrancesco et al. 2005); since the
available HXR measures of the Ophiuchus cluster have no sufficient
spatial resolution, we assume in the following that the radial
distribution function of the non-thermal electrons has the same
shape of the thermal gas radial distribution (see
Eq.\ref{eq.nth}). Under such assumption, we consistently assume
that the relativistic electrons extend out to the virial radius of
the cluster, like the IC plasma.\\
The value of the normalization of the electron spectrum $N_{e0}$
can be derived by reproducing the value of the HXR flux set by the
Swift-BAT and INTEGRAL experiments: for the previous spectral
index, we obtain the value $N_{e,0} = 1.1\times10^{-10}$
GeV$^{-1}$ cm$^{-3}$ for $p=3.5$, and $N_{e,0} =
2.1\times10^{-10}$ GeV$^{-1}$ cm$^{-3}$ for $p=4.4$.

All the previous information can be derived, strictly speaking,
only for the range of the electron energies that produce the HXR
emission via ICS. To obtain information on other energy ranges of
the electron spectrum one must consider other constraints.\\
The only other constraint that can be set on the electron spectrum
comes from the requirement that the heating rate of the IC gas
produced by non-thermal electron Coulomb collisions does not
exceed the bremsstrahlung cooling rate of the IC gas.
The heating rate produced by an electron with Lorentz
factor $\gamma$ and velocity $v=\beta c$ is given by
\begin{equation}
-{dE\over dt}\approx K \, z^2\, Z^2\,{1\over \beta}\,
   \left [\ln {2\, m_{\rm e}\, c^2\, \beta^2 \gamma^2\over I_{\rm p}}
 -\beta^2\right],
\label{dedx}
\end{equation}
where $Z^2$ is the (suitably averaged) squared charge of the
plasma's nuclei, $K\!=\!4\, \pi\, n_{th}\, r_{\rm e}^2\, m_{\rm
e}\, c^3$, with $r_{\rm e}\!=\!e^2/m_{\rm e}\, c^2\!\simeq\!2.82$
fm, and $I_{\rm p}\!=\!\hbar\, \omega_{\rm p}$, with $\omega_{\rm
p}\!=\! [4\pi\, n_{\rm e}\,e^2/m_{\rm e}]^{1/2}$ the plasma
frequency (see Colafrancesco et al. 2004, Colafrancesco \&
Marchegiani 2008).

As a consequence, the heating rate induced by the electrons with
the spectrum assumed in Eq.(\ref{spettroele}) is given by:
\begin{equation}
 {d \epsilon \over dt} \Biggr|_{_{\rm WR}} \equiv \int_{E_{min}}^{E_{max}} N_{e}(E,r)
 \bigg({dE \over dt}\bigg) dE \; ,
 \label{eq.heating}
\end{equation}
while the cooling rate is given by
\begin{eqnarray}
&&{d\epsilon\over dt}\Biggr|_{_{\rm X}}= a\, [n_{\rm th}(r)]^2\, \sqrt{kT(r,t)};\nonumber\\
&&a= \sqrt{2^{11}\pi^3\over 3^3}\; {e^6\sqrt{m_{\rm e} }\over h\,m_{\rm e}^2\, c^3}\,
{\bar G}\, {\bar z}\nonumber\\ &&~\sim 4.8\times 10^{-24}\,{\bar z}\,{1\over \sqrt{\rm
keV}} {\rm erg\,cm^3 \over s}\; ,
 \label{Xloss}
\end{eqnarray}
where $\bar z$ is an average charge of the IC plasma (we have approximated here the Gaunt
factor ${\bar G}$ by unity).

The two expressions in Eqs. (\ref{eq.heating}) and (\ref{Xloss})
equal (to the value $d\epsilon/dt \sim 4.7\times10^{-27}$ erg
cm$^{-3}$ s$^{-1}$) for the energy $E_{min}\sim 33$ MeV for
$p=3.5$, and for $E_{min}\sim 91$ MeV for $p=4.4$; therefore, we
assume these energy values as the minimum energy of the primary
electron spectra. As for the maximum electron energy we can safely
choose $E_{max}\rightarrow \infty$ since its specific value is
irrelevant for the assumed spectral index.\\
Under such assumptions, it is possible to calculate the overall radiation
emission via the various emission mechanisms in different frequency ranges.

The synchrotron emission spectrum produced at radio frequencies by
primary electrons is shown in Fig.\ref{fig.radioprim} for
different values of the magnetic field (the B-field has been
assumed to be constant in the emission region, and this
corresponds, approximately, to consider a volume averaged value of
the magnetic field). From the radio emission, we can derive a
value of the average magnetic field of $\sim$ 0.1 $\mu$G for
$p=3.5$ and $\sim$ 0.2 $\mu$G for $p=4.4$, in agreement with that
derived by Ajello et al. (2009) and Perez-Torres et al. (2008).

\begin{figure}[ht]
\begin{center}
\vbox{
 \epsfig{file=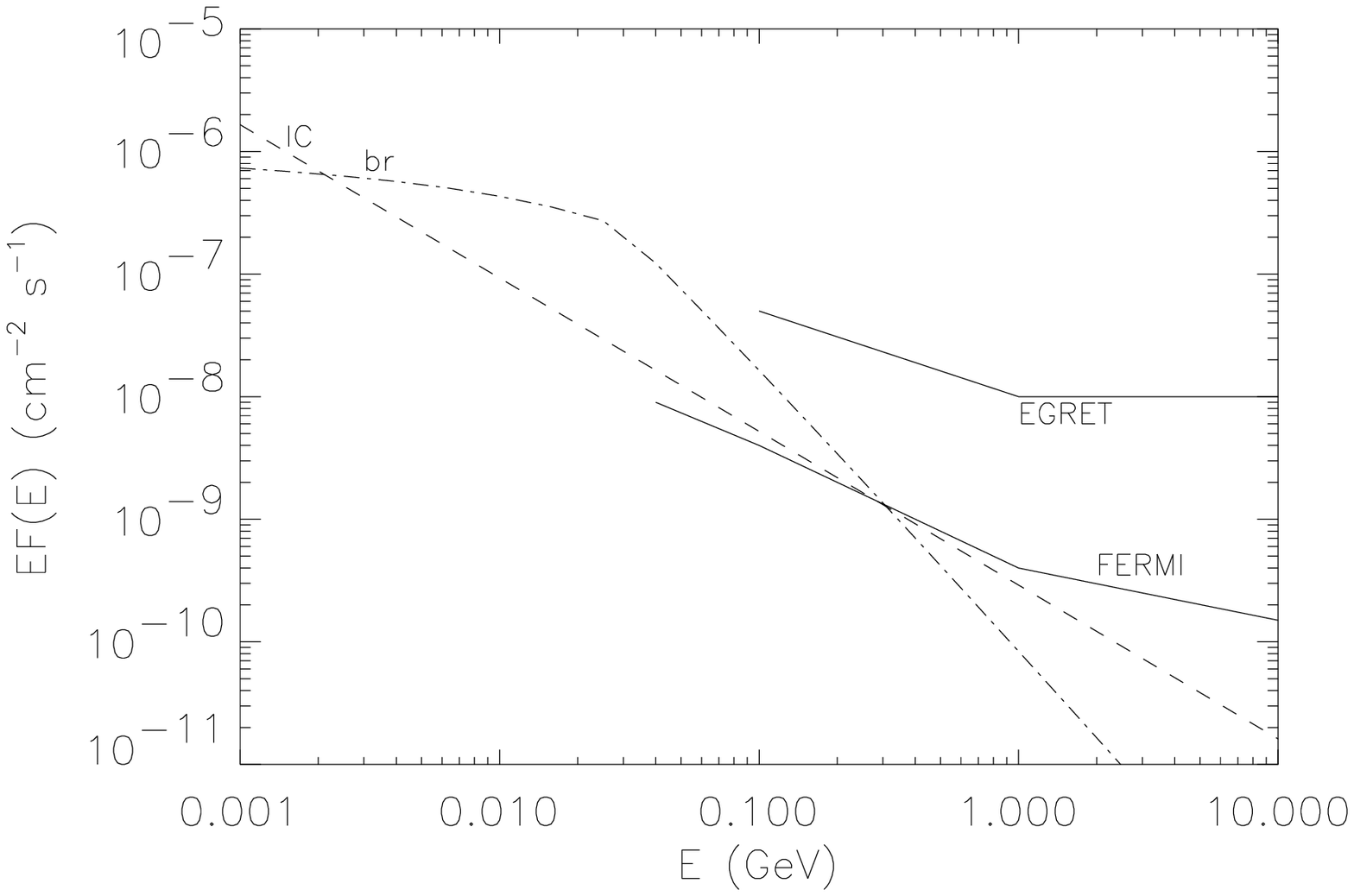,height=8.cm,width=8.cm,angle=0.0}
 \epsfig{file=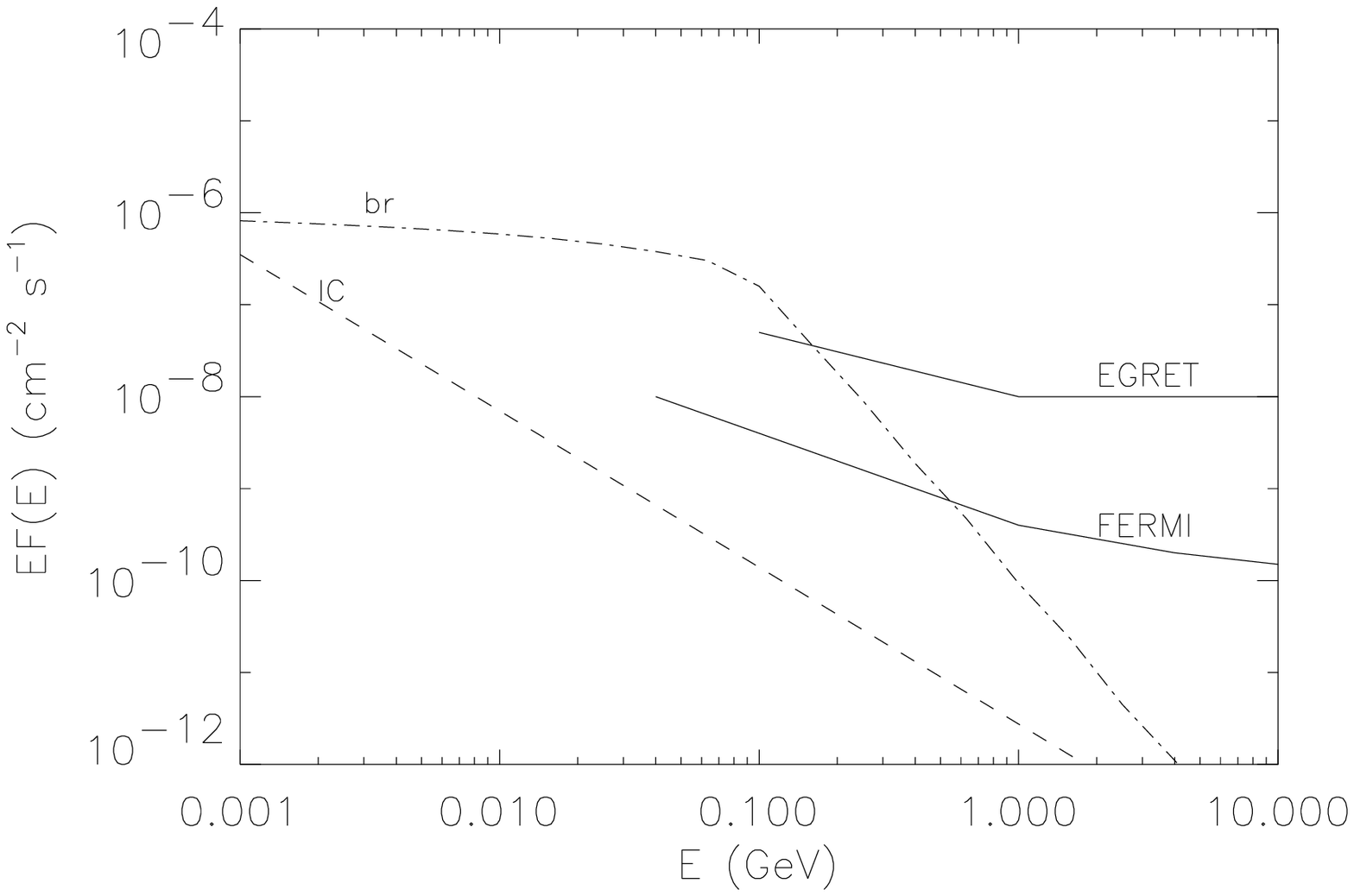,height=8.cm,width=8.cm,angle=0.0}
}
\end{center}
 \caption{\footnotesize{The gamma-ray spectrum produced by primary electrons
with $p=3.5$ (upper panel) and $p=4.4$ (lower panel) via ICS
 (dashed) and bremsstrahlung (dot-dashed) are compared to the sensitivity curves
 of EGRET and Fermi (5$\sigma$, 1 year observation).}}
 \label{fig.gammaprim}
\end{figure}

In the gamma-ray frequency range, these primary electrons emit via non-thermal
bremsstrahlung and ICS against CMB radiation field if their energy spectrum, as
we assume in this case, extends up to high energies (at least up to $E\sim
100-1000$ GeV in order to produce ICS emission in the energy range 0.1--10
GeV).
Fig. \ref{fig.gammaprim} shows that the EGRET upper limit on
Ophiuchus, $F(>100 \mbox{ MeV}) \leq 5\times10^{-8}$ cm$^{-2}$
s$^{-1}$ (Reimer et al. 2003), is not exceeded in the $p=3.5$
case, while it is marginally exceeded in the $p=4.4$ case.
In the first case, we can conclude that the HXR observation of
Ophiuchus cluster sets a constraint on the ICS emission from
relativistic electrons that is stronger than the analogous limit
set by EGRET; in the second case, the EGRET limit is stronger than
the HXR limit. The signals we derive here for the gamma-ray
emission of Ophiuchus in the PEM, and in particular the one
derived from non-thermal bremsstrahlung emission, are sensibly
larger than the Fermi sensitivity at $E \simlt 300$ MeV;
therefore, such an experiment could be able either to detect the
bremsstrahlung gamma-ray emission from Ophiuchus or set even
stronger limits on the non-thermal electron density.

It is important to stress here that the HXR data sets also
indirectly a lower cut-off of $E_{min}\sim 33$ and 90 MeV for the
two models considered on the electron spectrum in order to have an
heating rate not larger than the cooling rate.

\subsection{Secondary Electron Model from pp collisions (SEM-pp)}

To calculate the overall radiation emission from secondary
electrons produced by collisions of cosmic ray (CR) non-thermal
protons and the thermal protons of the IC gas (see Marchegiani et
al. 2007 for details), we assume that the non-thermal protons have
the following spectrum
\begin{equation}
 N_p(E,r)=N_{p,0} (E/\mbox{GeV})^{-s} g(r) \; ,
 \label{spettroproto}
\end{equation}
and we further assume, also in this case, that their spatial
distribution is the same of the thermal IC gas out to the virial
radius.\\
Similarly to the PEM model, we assume proton spectral indices
$s=2.5$ and $s=3.4$, which provide again radio spectral indices of
$\alpha_R=1.25$ and 1.7 respectively.

\begin{figure}[ht]
\begin{center}
\vbox{
 \epsfig{file=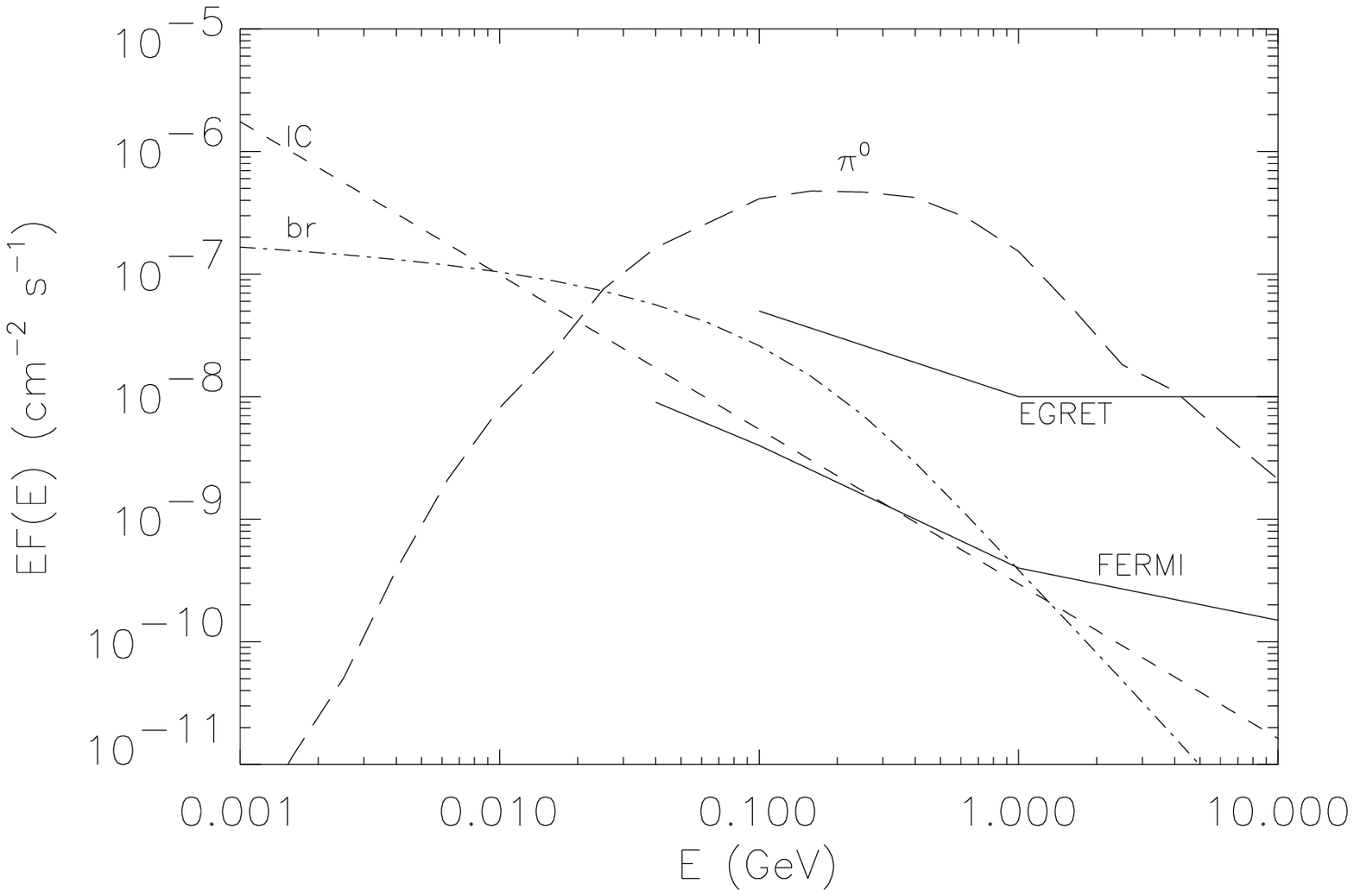,height=8.cm,width=8.cm,angle=0.0}
 \epsfig{file=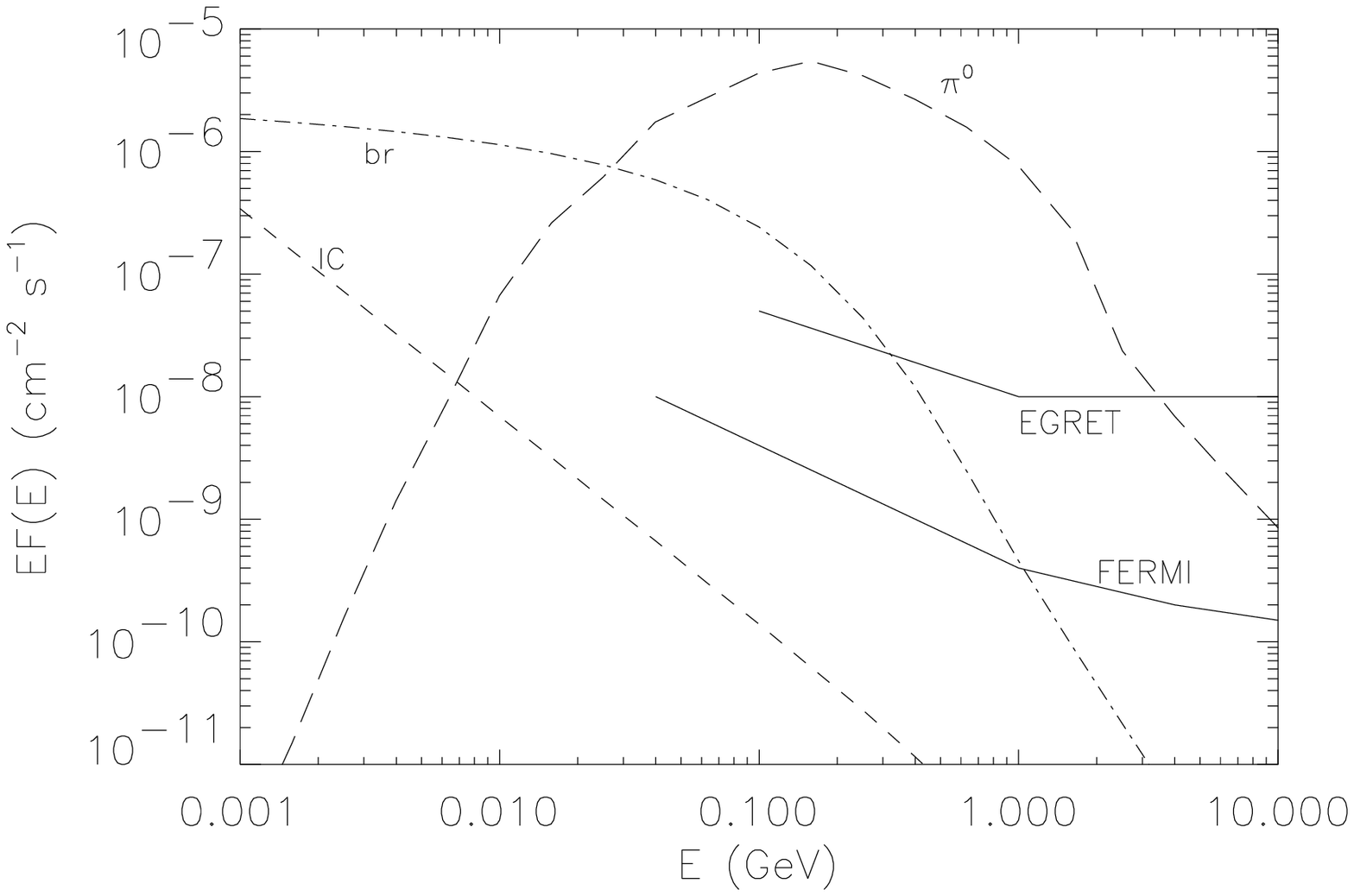,height=8.cm,width=8.cm,angle=0.0}
}
\end{center}
 \caption{\footnotesize{The gamma-ray spectrum of Ophiuchus
with $s=2.5$ (upper panel) and $s=3.4$ (lower panel) as produced by ICS
 (dashes) and bremsstrahlung (dot dashes) of secondary electrons and by neutral pion decay
 (long dashes).
 We compare the predictions of the SEM-pp model with the sensitivity curves of EGRET and
 Fermi (5$\sigma$, 1 year observation).
 }}
 \label{fig.gammasec}
\end{figure}

The HXR data explained in terms of the ICS emission of the
secondary electrons, provides values $N_{p,0}= 1.5\times10^{-6}$
GeV$^{-1}$ cm$^{-3}$ (for $s=2.5$) and $N_{p,0} =
3.4\times10^{-5}$ GeV$^{-1}$ cm$^{-3}$ (for $s=3.4$) for the
normalization of the spectrum in Eq.(\ref{spettroproto}).\\
Such values of the proton density normalization imply serious
problems for Ophiuchus:\\
i) first, the proton pressure at the center of the cluster is more
than 5 times larger than the thermal gas pressure for $s=2.5$ and
367 times larger for $s=3.4$, a fact that sets serious problems to
the cluster stability;\\
ii) secondly, the heating rate (Coulomb losses and hadronic
collisions) induced by non-thermal protons in the cluster center
is $d\epsilon/dt \sim 1.1\times10^{-25}$ erg cm$^{-3}$ s$^{-1}$
for $s=2.5$, and $d\epsilon/dt \sim 1.4\times10^{-24}$ erg
cm$^{-3}$ s$^{-1}$ for $s=3.4$; these values are about 23 and 298
times larger than the cooling rate. This fact would imply a quite
fast heating of the cluster, $d(kT)/dt \sim 41$ keV Gyr$^{-1}$ and
$d(kT)/dt \sim 531$ keV Gyr$^{-1}$, that will bring in a short
time the whole IC gas to a temperature sensitively different
(larger) from the observed one;\\
iii) finally, the gamma-ray emission produced by both secondary
electrons and by neutral pion decay (see Fig.\ref{fig.gammasec})
exceeds the EGRET upper limit on Ophiuchus by a factor $\sim$ 18
and 170, for $s=2.5$ and 3.4 respectively.

Thus, we must conclude that the HXR emission of Ophiuchus as set
by Swift and INTEGRAL, cannot be produced by secondary SEM-pp
electrons. We notice that such a conclusion is analogous to that
found in the case of other clusters we already studied like Coma,
A2199, A2163 and Perseus (Colafrancesco \& Marchegiani 2008).

\subsection{Secondary Electron Model from DM annihilation (SEM-DM)}

We consider here three neutralino DM models (similarly to the
analysis of Profumo 2008) with neutralino masses  $M_{\chi}= 81
(W^+W^-), 40(b {\bar b})$ and $10 (\tau^+ \tau^-)$ GeV.

For each neutralino model we consider a radial DM density profile as given by
\begin{equation}
 g_{DM}(r)=\exp [-(2/\alpha)((r/r_c)^\alpha-1)]
 \label{eq.prof.dm}
\end{equation}
(Navarro et al. 2004), with $\alpha=0.17$ and $r_c$ equal to the
core radius of the thermal gas density distribution. We assume
that this DM radial profile extends out to the virial radius.
The spectrum of the DM source function for the secondary electrons has,
consequently, a radial distribution $\propto g_{DM}^2(r)$.

To derive the equilibrium spectrum of these secondary electrons in
Ophiuchus we consider the role of the dominant energy loss
mechanisms. These are ICS losses against CMB photons and
synchrotron losses for electrons with energy larger than a few
hundreds MeV (notice that synchrotron losses for magnetic fields
less than 3 $\mu$G, are negligible with respect to the ICS
losses), while at low energies ($\simlt 150$ MeV) the dominant
energy loss mechanisms are Coulombian interactions with the IC gas
particles.\\
For this reason the final spatial distribution of secondary
electrons is proportional to $g_{DM}^2(r)$ at high energies ($>
150$ MeV) and proportional to $g_{DM}^2(r)/n_{th}(r)$ at low
energies ($<150$ MeV).

The DM-produced secondary electron density is fixed, also in this case, by
requiring that their ICS emission fits the observed HXR emission; such a
constraint corresponds to set the value of the neutralino annihilation cross
section, $\langle \sigma V \rangle$, because both the neutralino mass and its
composition have been fixed by the chosen model.

\begin{figure}[ht]
\begin{center}
 \epsfig{file=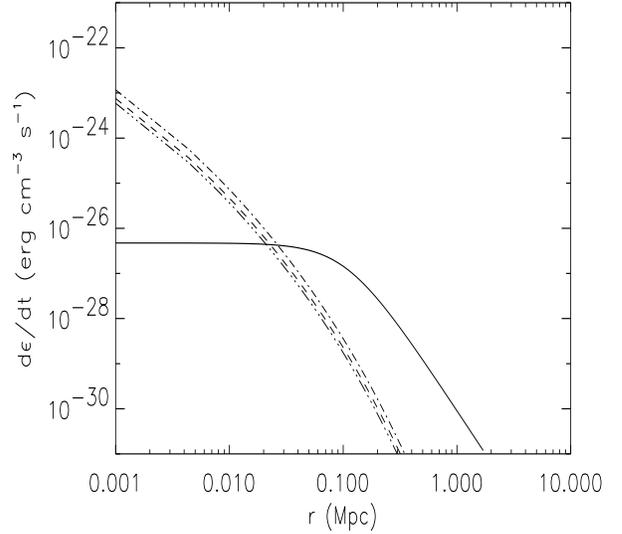,height=8.cm,width=8.cm,angle=0.0}
\end{center}
 \caption{\footnotesize{The heating rate induced by secondary electrons
 produced by DM annihilation is shown at different radii for different neutralino models:
$M_{\chi}=$ 81 GeV (dashed), 40 GeV (dot-dashed) and 10 GeV
(dot-dot dashed). The intracluster gas cooling rate (solid) is
also shown for comparison.
 }}
 \label{fig.heatingdm}
\end{figure}

\begin{figure}[ht]
\begin{center}
 \epsfig{file=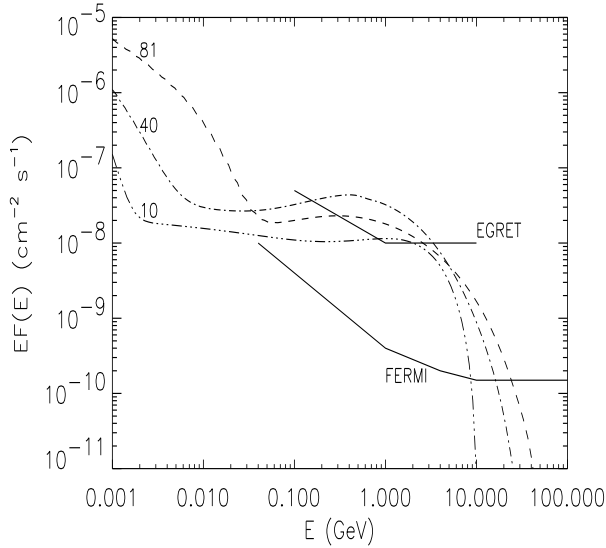,height=8.cm,width=8.cm,angle=0.0}
\end{center}
 \caption{\footnotesize{The overall gamma-ray spectrum produced by the
 composition of the contributions of the secondary SEM-DM electrons (ICS and bremsstrahlung)
 and of the neutral pion decay for the three neutralino models here considered.
 The sensitivity curves of EGRET and Fermi (5$\sigma$, 1 year observation) are shown
 for comparison. }}
 \label{fig.gammadm}
\end{figure}

\begin{figure}[ht]
\begin{center}
 \epsfig{file=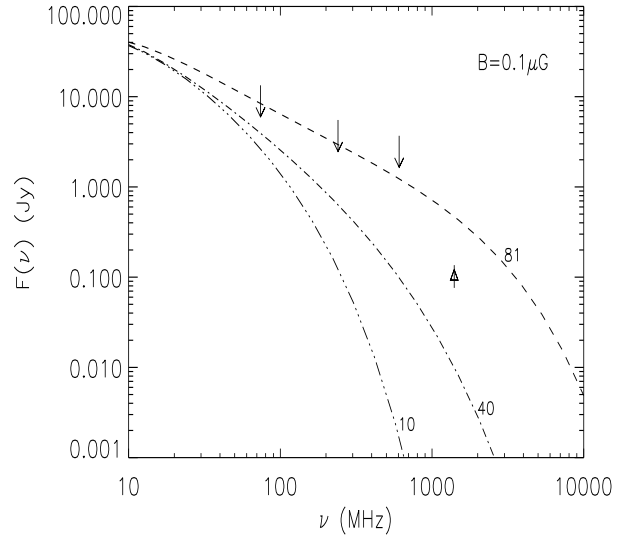,height=8.cm,width=8.cm,angle=0.0}
\end{center}
 \caption{\footnotesize{The radio halo spectrum produced from secondary SEM-DM
 electrons via synchrotron emission in a constant magnetic field of 0.1 $\mu$G.
Data are from Perez-Torres et al. (2008) (upper limits) and Govoni
et al. (2009) (point at 1.4 GHz). }}
 \label{fig.radiodm}
\end{figure}

\begin{figure}[ht]
\begin{center}
 \epsfig{file=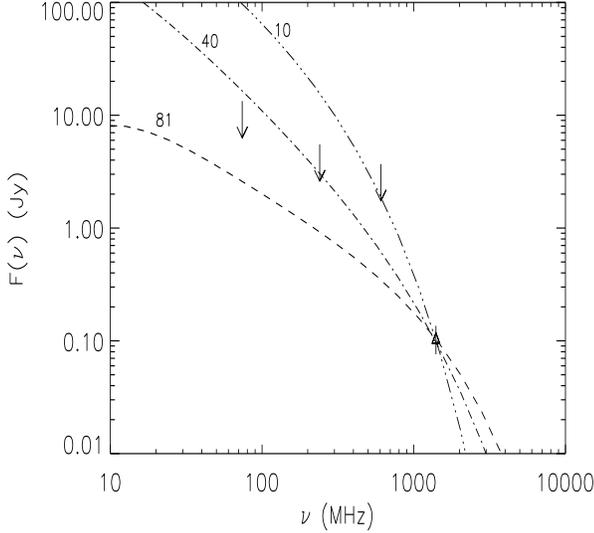,height=8.cm,width=8.cm,angle=0.0}
\end{center}
 \caption{\footnotesize{The radio halo spectrum produced from secondary SEM-DM
 electrons via synchrotron emission in a constant magnetic field of 0.055, 0.18
and 0.39 $\mu$G for $M_\chi=81$, 40 and 10 GeV, respectively.
Data are from Perez-Torres et al. (2008) (upper limits)
and Govoni et al. (2009) (point at 1.4 GHz).
}}
 \label{fig.radiodm2}
\end{figure}

Also this SEM-DM model has serious implications for the Ophiuchus
cluster.\\
i) The heating rate at the cluster center as produced by the
secondary SEM-DM electrons is very high; Fig. \ref{fig.heatingdm}
shows the secondary electrons heating rate at different radii
compared to the cooling rate of the thermal IC gas. In fact, the
heating rate largely exceeds the cooling rate in the cluster core
at $r<30$ kpc. This result would imply a fast over-heating of the
Ophiuchus core, even though the volume integral of the heating
rate is always lower than the volume integral of the cooling rate
(this last quantity is $1.5\times10^{44}$ erg/s, while the
integrate heating rate is $4.6\times10^{42}$ erg/s,
$7.4\times10^{42}$ erg/s and $3.6\times10^{42}$ erg/s for
$M_\chi=$ 81, 40 and 10 GeV, respectively).

ii) We show in Fig. \ref{fig.gammadm} the gamma-ray emission
spectra as produced by the DM composite model worked out here via
the three main mechanisms of gamma-ray emission: ICS and
bremsstrahlung from secondary electrons and neutral pion decay.
All the three DM models considered in this composite DM model for
Ophiuchus produce a gamma-ray flux that exceeds the EGRET limit,
$F(>100 \mbox{ MeV})=5.0\times10^{-8}$ cm$^{-2}$ s$^{-1}$ (the low
mass neutralino model with $M_{\chi}=10$ GeV is marginally
consistent with the EGRET limit). The gamma-ray flux of Ophiuchus
at $E>100$ MeV produced under the assumption that the same DM
model reproduce the HXR data are $7.4\times10^{-8}$,
$1.3\times10^{-7}$ and $4.3\times10^{-8}$ cm$^{-2}$ s$^{-1}$ for
neutralino masses of 81, 40 and 10 GeV.
A direct prediction of this DM model is that the gamma-ray flux produced by the
three neutralino models considered here should be easily detectable by the
Fermi experiment whose results will be able, therefore, to validate or rule out
this model for the origin of the HXR emission of Ophiuchus.

iii) Fig. \ref{fig.radiodm} shows the diffuse synchrotron radio
spectra produced by the same secondary SEM-DM electrons under the
assumption of a reference value of the average intracluster
magnetic field in Ophiuchus of 0.1 $\mu$G. This figure shows that,
for this value of magnetic field, a model whith neutralino mass
between 40 and 80 GeV can reproduce the radio data.\\
We also searched for the value of the magnetic field that, for
each of the models we considered, reproduces the Ophiuchus radio
halo flux at 1.4 GHz. Fig.\ref{fig.radiodm2} shows the radio
spectrum produced by SEM-DM electrons for best-fit magnetic field
values of 0.055, 0.18 and 0.39 $\mu$G, for $M_\chi=$ 81, 40 and 10
GeV, respectively. We can conclude that the 81 GeV ($W^+W^-$)
model is consistent with radio data, while the 10 GeV ($\tau^+
\tau^-$) model is not consistent. The 40 GeV ($b \bar{b}$) model
is a border-line situation: we find, in fact, that for a slightly
lower magnetic field of 0.17 $\mu$G the radio spectrum is
marginally consistent with the point at 1.4 GHz and the upper
limit at 74 MHz.

\section{Warming Ray Model}

In this Section we abandon the strategy of fitting the ICS
emission produced by high-E electrons to the HXR data of Swift and
INTEGRAL because we evaluate -- in the framework of the Warming
Ray (WR) model (see Colafrancesco \& Marchegiani 2008,
Colafrancesco et al. 2004) -- the spectral and spatial
characteristics of the WR proton population that produce through
their heating action the temperature structure of the Ophiuchus
cluster, namely a constant temperature radial profile at the
observed value of $kT \approx 9.9$ keV.

The proton spectrum is written as in Eq. (\ref{spettroproto}),
with the values $s=2.5$ and $s=3.4$, and assuming a radial
distribution given by $g(r)\propto g_{th}^\alpha(r)$, where the
value of $\alpha$ is found by fitting the radial temperature
profile of the cluster (see Colafrancesco \& Marchegiani 2008 for
technical details). The best fit analysis of the temperature
profile of Ophiuchus provides the value $\alpha=1$, in analogy to
what is found for other isothermal clusters (see discussion in
Colafrancesco \& Marchegiani 2008), while the central WR density
is $N_{p,0}=4.9\times10^{-8}$ cm$^{-3}$ GeV$^{-1}$ and
$N_{p,0}=9.4\times10^{-8}$ cm$^{-3}$ GeV$^{-1}$ for $s=2.5$ and
$s=3.4$ respectively, i.e. a factor $\approx 31$ and 362 lower
than that required to reproduce, in this model, the Swift
BAT/INTEGRAL HXR data.
Consequently, the pressure ratio of the WR to the thermal gas at
the cluster center is $P_{CR}/P_{th}\sim0.17$ for $s=2.5$ and
$P_{CR}/P_{th}\sim1.0$ for $s=3.4$: the first value does not give
any problem to the overall stability of the cluster, while the
second one is a problematic situation.

\begin{figure}[ht]
\begin{center}
\vbox{
 \epsfig{file=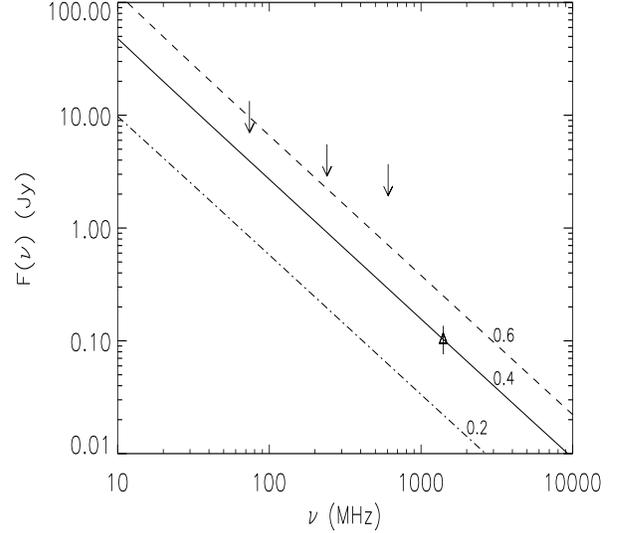,height=8.cm,width=8.cm,angle=0.0}
 \epsfig{file=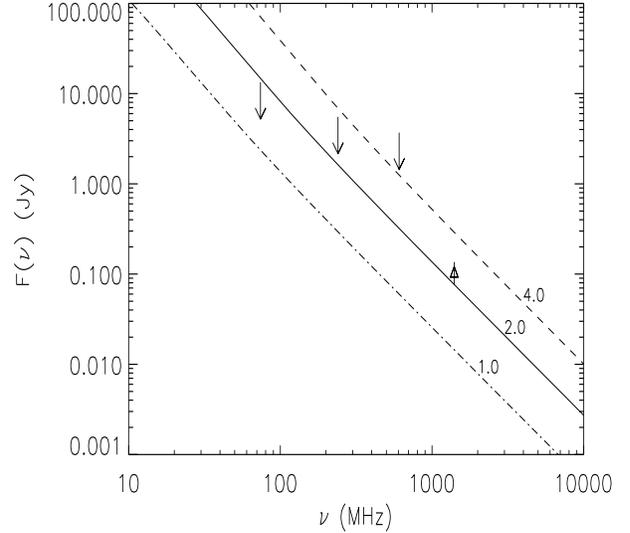,height=8.cm,width=8.cm,angle=0.0}
}
\end{center}
 \caption{\footnotesize{The radio halo spectrum produced by secondary electrons
 with $s=2.5$ (upper panel) and $s=3.4$ (lower panel)
 in the WR model is shown for different values of a constant IC magnetic field (as
 labeled) in units of $\mu$G.
Data are from Perez-Torres et al. (2008) (upper limits) and Govoni
et al. (2009) (point at 1.4 GHz). }}
 \label{fig.radioseccf}
\end{figure}

The diffuse radio emission produced by the secondary electron in
this WR model is shown in Fig. \ref{fig.radioseccf} for various
values of the uniform magnetic field; our results indicate that a
uniform B-field of the order of $\sim$ 0.4 $\mu$G is required to
fit the available data for $s=2.5$, and $B\sim2.0$ $\mu$G for
$s=3.4$.

\begin{figure}[ht]
\begin{center}
\vbox{
 \epsfig{file=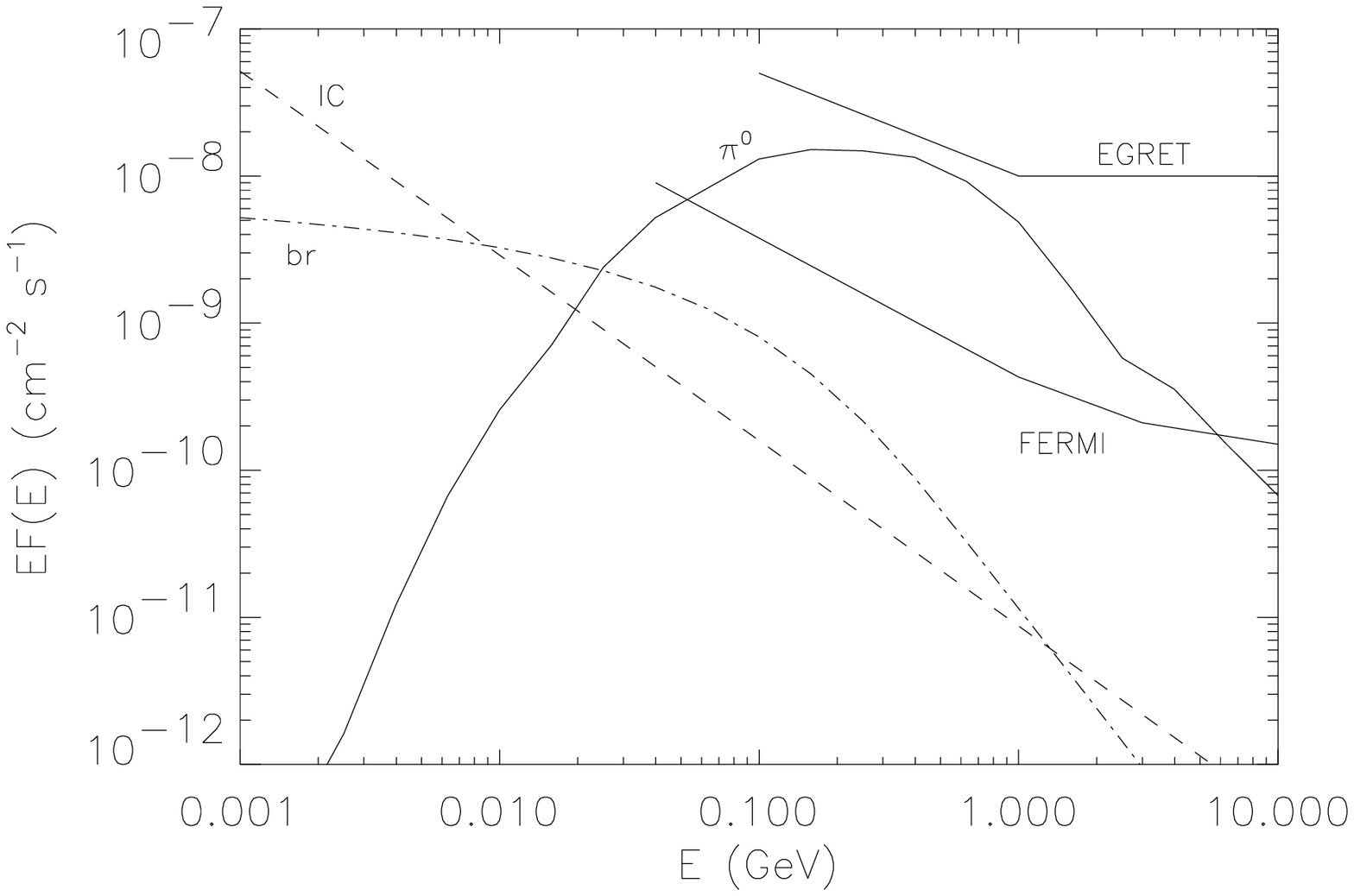,height=8.cm,width=8.cm,angle=0.0}
 \epsfig{file=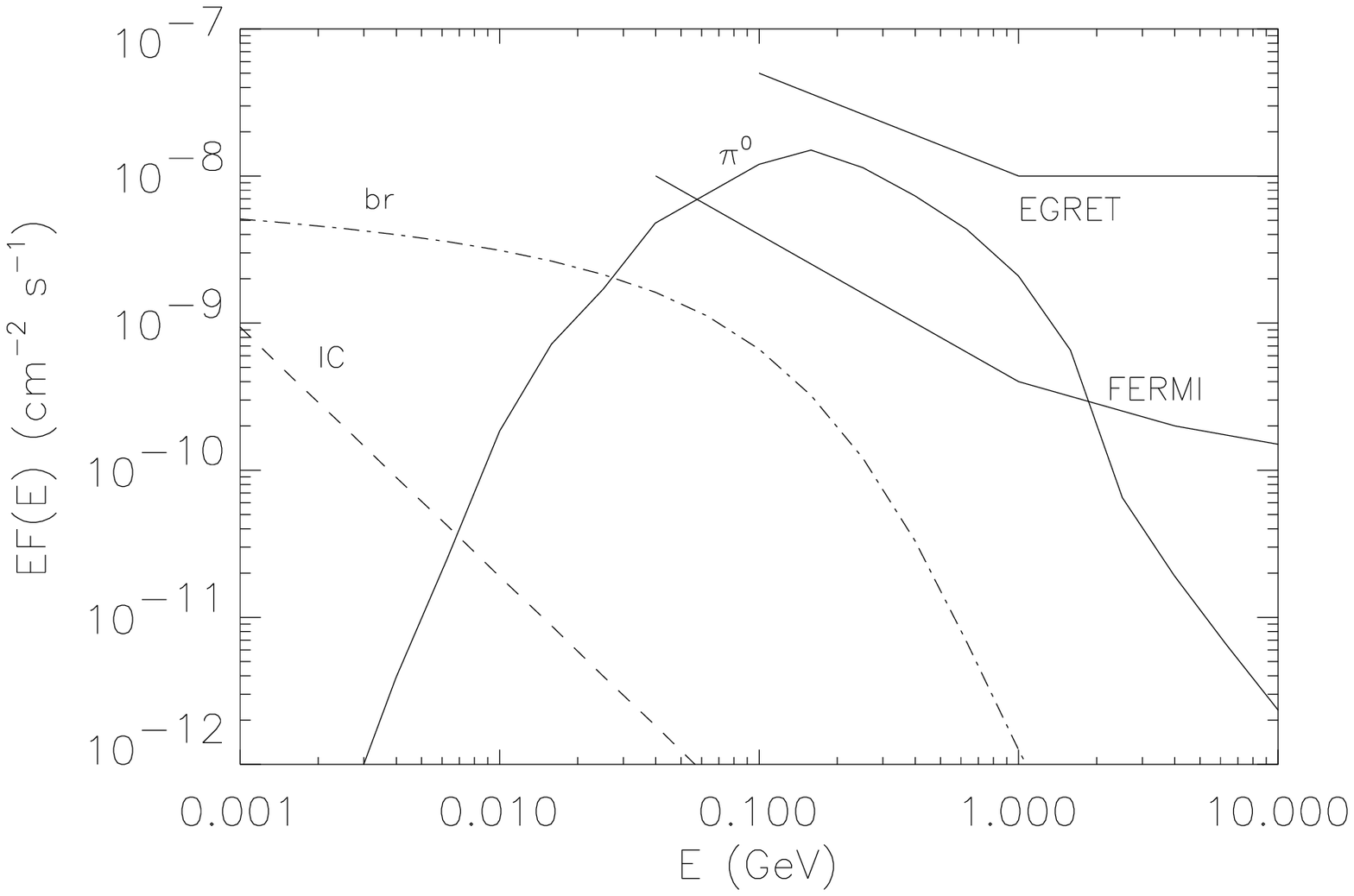,height=8.cm,width=8.cm,angle=0.0}
}
\end{center}
 \caption{\footnotesize{The gamma-ray spectrum produced by secondary electrons
 with $s=2.5$ (upper panel) and $s=3.4$ (lower panel)
 via ICS (dashes) and bremsstrahlung (dot-dashes) and neutral pion decay (solid)
 in the WR model that is set up by the requirement to reproduce the observed temperature
 structures of the IC gas in Ophiuchus.
 The sensitivity curves of EGRET and Fermi (5$\sigma$, 1 year observation) are shown
 for comparison.}}
 \label{fig.gammaseccf}
\end{figure}

Fig. \ref{fig.gammaseccf} shows the diffuse gamma-ray emission
from Ophiuchus as expected in the WR model: this emission consists
of the combination of the neutral pion decay spectrum and ICS and
bremsstrahlung emission from secondary electrons. The overall
gamma-ray emission of Ophiuchus in the WR model, $F(>100 \mbox{
MeV})=3.3\times10^{-8}$ cm$^{-2}$ s$^{-1}$ and $2.3\times10^{-8}$
cm$^{-2}$ s$^{-1}$ for $s=2.5$ and $s=3.4$ respectively, is below
the EGRET limit. However, the neutral pion decay gamma-ray
emission predicted in this WR model should be detectable by Fermi
in 1 yr observation (5$ \sigma$ c.l.).

\section{Discussion and conclusions}
 \label{sec.conclusions}

We have show in this paper that various and serious problems stand
out with the ICS interpretation of the HXR excess emission of the
Ophiuchus cluster and, in general, of galaxy clusters for which an
HXR emission excess has been detected. These problems are:

i) the actual level of the HXR emission flux: the derivation of an HXR emission
excess in clusters seems to depend strongly on the precise determination of the
background thermal bremsstrahlung emission.
Eckert et al. (2008) derived from INTEGRAL data an HXR flux in the 20--60 keV
band of $F_{HXR}=(10.1\pm2.5)\times10^{-12}$ erg cm$^{-2}$ s$^{-1}$, with an IC
gas temperature of $8.56^{+0.37}_{-0.35}$ keV.
Ajello et al. (2009), using Swift BAT data, derived an upper limit
($90 \%$ c.l.) $F_{HXR}\leq 7.2 \times10^{-12}$ erg cm$^{-2}$
s$^{-1}$, with a different IC gas temperature of
$9.5^{+1.4}_{-1.1}$ keV; it must be noticed that the same authors
(Ajello et al. 2009) derived the upper limit for the HXR flux of
$F_{HXR}\leq 4.5 \times10^{-12}$ erg cm$^{-2}$ s$^{-1}$ by using
an higher value of the temperature $9.93^{+0.24}_{-0.24}$ keV, as
obtained by using a combination of Chandra and Swift-BAT data.\\
In conclusion, it is clear that a crucial input quantity to
determine the value of the HXR excess flux is the detailed
modeling of the thermal emission of the IC gas, because different
values assumed for the IC gas temperature lead to different
conclusions on the amount of the HXR excess flux (see, e.g. the
long standing discussion on the evidence and counter evidence of
the HXR excess in Coma, Fusco-Femiano et al. 1999, 2004, 2007, 2008; Rossetti
\& Molendi 2004, 2007; see also Petrosian et al. 2008 for a
review).
For this reason, it would be extremely important to estimate the
temperature of the IC gas through measurements that are
independent from those obtained in the X-ray band. We notice, in
this context, that a reliable method to measure IC gas
temperatures can be found by using Sunyaev-Zel'dovich observations
over a wide spectral band (from $\sim 100$ to $\sim 400$ GHz)
reaching high frequencies where the sensitivity of the SZE to the
cluster temperature is maximum (we have discussed in details this
issue in a dedicated paper, Colafrancesco \& Marchegiani 2009; see
also Colafrancesco et al. 2003 and Colafrancesco 2007 for a
review);

ii) the ICS HXR scenario: the hypothesis that high-E electrons are
responsible for an ICS HXR emission at the level found by the
combination of Swift-BAT and INTEGRAL observations leads to some
important consequences.\\
First, in order to reconcile the HXR excess value with the
relative diffuse synchrotron radio emission (from the same
electron population) at the level observed in the same cluster,
the value of the average magnetic field must be quite low and of
the order of $\sim$ 0.1 $\mu$G for $p=3.5$ and $\sim 0.2$ $\mu$G
for $p=4.4$ (see Fig. \ref{fig.radioprim}). The result found for
Ophiuchus is analogous to what is derived for other clusters (see
also our previous results discussed in Colafrancesco, Marchegiani
\& Perola 2005, Marchegiani, Perola \& Colafrancesco 2007):
Specifically, we found that the data are consistent with an IC
magnetic field of order of $\sim 0.7$ and 1.2 $\mu$G at the
cluster center with a decreasing radial profile similar to that of
the IC gas, for $p=3.5$ and 4.4 respectively.\\
Secondly, there is a strong relation between the ICS HXR emission
level and the relative gamma-ray emission and the consequences on
the physics of the cluster, i.e. the heating of the IC gas and the
ratio between non-thermal and thermal pressures:\\
- if the electrons that produce the HXR emission are primaries,
their gamma-ray emission (dominated at $E < 1$ GeV by non-thermal
bremsstrahlung) is slightly lower than the EGRET upper limit in
the $p=3.5$ model, and slightly higher than this limit in the
$p=4.4$ model, but certainly detectable by Fermi (see
Fig.\ref{fig.gammaprim}). If Fermi will not detect such gamma-ray
emission, one should conclude that the ICS HXR emission is much
lower than the Swift-INTEGRAL HXR detection and that the
relativistic electron content of Ophiuchus is consequently much
lower.
The HXR data set, in addition, a lower cut-off of $E_{min}\sim 33$
and 93 MeV (for $p=$3.5 and 4.4) on the electron spectrum in order
to have an heating rate not larger than the cooling rate;\\
- if the electrons responsible for the ICS HXR emission are
secondary particles produced in the decay of charged pions
generated by cosmic-ray proton collisions with the IC gas protons
(SEM-pp), then an ICS flux set at the HXR observations has
unacceptable consequences.
Specifically we find that: in the $s=2.5$ case, the pressure
exerted by relativistic protons at the cluster center is $\sim 5$
times larger than the thermal gas one; the heating rate induced by
the same relativistic protons at the cluster center is $\sim 23$
times larger than the IC gas cooling rate; and the gamma-ray
emission produced by neutral pion decay exceeds the EGRET limit by
a factor $\sim 18$; in the $s=3.4$ case, these quantities rise
respectively to $\sim$ 367, 298 and 170 (see
Fig.\ref{fig.gammasec}). For all these reasons, we conclude that
if electrons produce an ICS HXR emission in the observed range,
they cannot be secondary (in the SEM-pp). This conclusion is
analogous to what has been found also in other clusters like Coma,
A2199, A2163 and Perseus (see Colafrancesco \& Marchegiani
2008);\\
- if electrons are produced by neutralino DM annihilation, we have
found that: the heating rate they induce at the cluster center is
quite high (see Fig. \ref{fig.heatingdm}); the relative gamma-ray
emission exceeds the EGRET limit for the two high-$M_{\chi}$
models here considered ($M_{\chi} = 40$ and $81$ GeV) (see Fig.
\ref{fig.gammadm}), with a marginal consistency for the low-mass
model with $M_\chi=10$ GeV; the radio flux produced by electrons
is consistent with the available data for $M_\chi>40$ GeV, and for
$B<0.18$ $\mu$G. Therefore, the information inferred by gamma and
radio data are not compatible, and we conclude that it is not
possible to conceive that the ICS HXR emission of secondary SEM-DM
electrons has a flux close to the available observation (i.e., the
maximum allowed flux set by Swift and INTEGRAL, see Sect.1), and
thus their annihilation cross section must be much lower than the
values used by Profumo (2008).\\
Even normalizing these models to the lower allowed flux value of
the HXR excess of Ophiuchus (see our discussion in Sect.1), all
the previous results vary (decrease) by $\sim 15 \%$, leaving
unchanged our basic conclusions.

iii) Relaxing the assumption to recover the observed HXR excess
and assuming that non-thermal protons act as warming rays (see
Colafrancesco \& Marchegiani 2008) it is possible to paint a much
more acceptable scenario in which the unacceptable pressure ratios
derived in SEM models do not hold since the ratio
$P_{non-th}/P_{th} \approx 0.17$ and 1.0 for, respectively,
$s=2.5$ and $3.4$ and it is constant throughout all the cluster
(this is because non-thermal protons must have the same spatial
distribution of the thermal IC gas to recover the spatial
temperature distribution of the cluster).
In addition, the WR model has other positive aspects for the
cluster structure:
i) it does not induce excess heating effects, since a
quasi-stationary balance between heating and cooling is the
working assumption of the WR model; ii) we found that the diffuse
radio emission produced in this case requires, for $s=2.5$ and
$s=3.4$ respectively, a value of the average magnetic field of
$\sim$ 0.4 and 2 $\mu$G (see Fig. \ref{fig.radioseccf}) and a
central value of $\sim$ 1.1 and 6 $\mu$G with a radial profile
similar to that of the IC gas, consistently with the general
findings for clusters through Faraday Rotation measurements (see,
e.g., Carilli \& Taylor 2002, Govoni \& Feretti 2004); iii) the
gamma-ray emission produced in this model is quite lower than the
EGRET limit but definitely detectable by Fermi (see Fig.
\ref{fig.gammaseccf}). The Fermi detection of such gamma-ray
emission from Ophiuchus will have a crucial impact for proving or
disproving this model.\\
In such a WR model, the HXR ICS flux of Ophiuchus is much lower
(by a factor $\sim 30$ and 362 for $s=2.5$ and $s=3.4$
respectively) than the limit set by the present observations (by
INTEGRAL and Swift-BAT) and could only be detectable by using long
exposure observations with the next generation HXR instruments
like NeXT (see e.g. Takahashi et al. 2004).

To conclude, models of high-E electrons in clusters that can be adjusted to
reproduce their ICS HXR emission at the level indicated by the available
observations fail to work because they would imply unacceptable levels of
heating and gamma-ray emission.
On the contrary, models of high-energy particles that are able to
reproduce the IC gas temperature structure (i.e. that WR model)
predict a level of non-thermal HXR ICS emission that is far below
the current limits obtained with INTEGRAL and Swift-BAT, and
provide an overall Spectral Energy Distribution that is consistent
with all the available data -- from radio to gamma-rays -- on
Ophiuchus as well as on other clusters.

\begin{acknowledgements}
We thank P.Ullio for providing detailed source spectra of the
neutralino annihilation models considered in this paper.
We also thank the Referee, M.A. Perez-Torres, for his comments and
suggestions that allowed to improve the presentation of our
results.
\end{acknowledgements}


\end{document}